\documentclass[aps,twocolumn,showpacs]{revtex4}
\usepackage{epsfig}
\def\avg#1{\langle#1\rangle}
\def\Re{\rm{Re}}
\def\Im{\rm{Im}}
\def\be{\begin{equation}}       \def\ee{\end{equation}}
\def\bea{\begin{eqnarray}}      \def\eea{\end{eqnarray}}
\def\pp{\parallel}

\begin{document}
\title{A sufficient condition for the absence of the sign problem in
the fermionic quantum Monte-Carlo algorithm}
\author{Congjun Wu and Shou-Cheng Zhang}
\affiliation{Department of Physics, McCullough Building, Stanford
University, Stanford CA 94305-4045}

\begin{abstract}
Quantum Monte-Carlo (QMC) simulations involving fermions have the
notorious sign problem. Some well-known exceptions of the
auxiliary field QMC algorithm rely on the factorizibility of the
fermion determinant.
Recently, a fermionic QMC algorithm\cite{WU2003} has been found in
which the fermion determinant may not necessarily factorizable,
but can instead be expressed as a product of complex conjugate
pairs of eigenvalues, thus eliminating the sign problem for a much
wider class of models. In this paper, we present general
conditions for the applicability of this algorithm and point out
that it is deeply related to the time reversal symmetry of the
fermion matrix. We apply this method to various models of strongly
correlated systems at all doping levels and lattice geometries,
and show that many novel phases can be simulated without the sign
problem.
\end{abstract}
\maketitle

\section{Introduction}
Understanding the physics of strongly correlated many body systems
is a main focus of condensed matter physics today. However, most
models with strong interactions can not be solved exactly except
in one dimension. Presently, there are no systematic
non-perturbative analytic methods which work in higher dimensions.
Largely because of this reason, numerical simulations such as
exact diagonalization (ED), density-matrix renormalization group
(DMRG), quantum Monte-Carlo (QMC) are extensively performed to
study strongly correlated systems. However, each of the numerical
methods has its own limitations. The ED can only be
performed in a very small sample size, and the DMRG method is
largely restricted to one-dimensional systems. In contrast, QMC
simulation is the only systematic and scalable method with
sufficient numerical accuracy for higher dimensional problems.
However, QMC also has the notorious fermion sign problem which
makes low temperature properties inaccessible.

In lattice systems, a particular version of QMC uses the
auxiliary field method introduced by Blankenbecler, Scalapino and
Sugar \cite{BLANKENBECLER1981}, with fruitful results. Because one
can not directly sample the fermionic Grassmann fields, the
standard process is to perform a Hubbard-Stratonovich (HS)
transformation to decouple the four-fermion interaction terms, and
then to integrate out the fermions \cite{BLANKENBECLER1981}. The
resulting fermion functional determinant works as the statistical
weight for sampling the auxiliary fields. However, generally
speaking the fermion determinant may not be positive, and can even
be complex in some cases. The sign or the phase of the fermion
determinants can lead to dramatic cancellations which makes
statistical errors to scale exponentially as the inverse of the
temperature and size of the system. This notorious sign problem is
the major obstacle in applying QMC to fermionic systems. A
successful solution to the sign problem would obviously lead to
great advances in quantum many body physics.

There are a few exceptions where the sign problem is absent, such
as the negative $U$ Hubbard model and the positive $U$ Hubbard
model in a bipartite lattice at the half-filling \cite{HIRSCH1985}.
In both cases, the fermion determinant after the HS decomposition
can be factorized into to two real parts with the same
sign. It is therefore positive-definite. Unfortunately, general fermion
determinants may not be factorizable for more complicated models
and the majority of models do have the sign problem. In recent
years, several other algorithms have been proposed which partially
solves the minus sign problem
\cite{chandrasekharan1999, koonin1997, rombouts1998, imada2000, kashima2001}.

Recently, it has been shown that the minus sign problem can be
eliminated without relying on the factorizibility of the fermion
determinant, therefore, a broader class of models can be simulated
by the QMC algorithm\cite{WU2003}. The fermion determinant can
always be expressed as a product of its eigenvalues; under certain
conditions, the eigenvalues of the fermion determinant always
appear in complex conjugate pairs, thus making the fermion
determinant positive definite. In this article, we shall show that
the property of conjugate eigenvalue pairs follows from the time
reversal symmetry of the HS decoupled Hamiltonian, and can be
viewed as a generalization of the Kramers theorem in quantum
mechanics. We shall call this method the {\it $T$-invariant
decomposition} (time reversal invariant decomposition). This
method does not lead to any improvement for the single band
Hubbard model, but significantly extends the applicability of the
QMC to multi-band, multi-layer or higher spin models. This
algorithm is particularly useful for Hubbard models with higher
spins, which can be accurately realized in systems of cold atoms.
Recently, Assaad {\it et al}. \cite{ASSAAD2003} applied the QMC to
generalized Hubbard models with more bands. Imposing the
factorizibility condition of the fermion determinant, they found
that they could extend the parameter regime for QMC free of the
sign problem only by scarifying the spin rotational invariance.
However, applying our method of $T$-invariant decomposition
without requiring factorizibility, we shall show that multi-band
or higher spin Hubbard models can be simulated for an extended
parameter regime without scarifying the spin rotational
invariance. This QMC algorithm based on $T$-invariant
decomposition has been recently applied to conclusively
demonstrate the staggered current carrying ground state in a
bi-layer model \cite{CAPPONI2004}.

The rest of this article is outlined as follows: In Sec.
\ref{sect:spin12}, the sign problem for the spin 1/2 Hubbard model
is reviewed. In Sec. \ref{sect:theorem}, we prove the fundamental
theorem of $T$-invariant decomposition and show the absence of the
sign problem. In Sec. \ref{sect:spin32}, we employ the algorithm
to the spin 3/2 Hubbard model and the generalized arbitrary spin
$n-1/2$ fermionic Hubbard model. In Sec. \ref{sect:SZH}, we apply
it to a bi-layer model introduced by Scalapino, Zhang and
Hanke\cite{SCALAPINO1998}, which can be mapped into the spin 3/2
Hubbard model. In Sec. \ref{sect:exotic}, we discuss the algorithm
in the model Hamiltonians with bond interactions and various
exotic phases. Final conclusions are presented in Sec.
\ref{sect:conclusion}.

\section{The sign problem in the spin 1/2 Hubbard model}\label{sect:spin12}
In this section, we review the sign problem in the spin 1/2
Hubbard model and interpret its absence in the negative $U$ case
as due to its time reversal properties of the HS decomposition.
The Hubbard model on the lattice is commonly defined as
\bea\label{hubbard12} H&=&-t\sum_{ij,\sigma}(c^\dagger_{i\sigma}
c_{j\sigma} +h.c.)
-\mu \sum_i n(i)\nonumber \\
&+&U \sum_i (n_\uparrow(i)-\frac{1}{2}) (
n_\downarrow(i)-\frac{1}{2}), \eea with $t$ the hopping integral,
$\mu$ the chemical potential, $\sigma=\uparrow,\downarrow$,
$n_\sigma(i)=c^\dagger_{i\sigma} c_{i\sigma}$ and $n(i)=
n_\uparrow(i)+n_\downarrow(i)$. At half-filling and on a bipartite
lattice, the particle-hole symmetry ensures that $\mu=0$.

To perform  the QMC simulation, we first need to decouple the
4-fermion interaction terms using the HS transformations by the
Gaussian integral: \bea \exp( \frac{1}{2} A^2)=\sqrt{2\pi} \int dx
\exp (-\frac{1}{2} x^2- x A),
\nonumber \\
\exp(-\frac{1}{2} A^2)=\sqrt{2\pi} \int dx \exp (-\frac{1}{2} x^2-
i x A) . \eea Various HS decoupling schemes are discussed in Ref.
\cite{HIRSCH1983}. For $U<0$, it is convenient to decouple Eq.
(\ref{hubbard12}) in the density channel and then integrate out the
fermions. The resulting partition function is given by
\begin{widetext}
\bea
Z&=&\int D c^\dagger D c ~~ \exp \{ -\int^\beta_0 d
\tau ~~(c^\dagger_\sigma \frac{\partial}{\partial \tau}
c_\sigma +H) \} \nonumber \\
&=&\int D n D c^\dagger D c ~ \exp \Big\{ -\frac{|U|}{2}
\int^\beta_0 d\tau  \sum_i(n(i,\tau)-1)^2\Big\}
\exp\Big \{-\int^\beta_0~ d\tau ~(H_K+H_I(\tau)) \Big\}
\nonumber \\
&=&\int D n \exp \Big\{ -\frac{|U|}{2}
\int^\beta_0 d\tau  \sum_i(n(i,\tau)-1)^2
\Big\} ~ \det \{  I+B \},
\eea
\end{widetext}
where $n(i,\tau)$ is a real HS bose density field.
The imaginary-time independent kinetic energy term $H_K$ and the
imaginary-time dependent decoupled interaction term $H_I(\tau)$
can be expressed as
\bea\label{density-HS}
H_K&=& \sum_{ij} c^\dagger_{i\sigma} h^K_{ij,\sigma \sigma^\prime}
c_{j\sigma^\prime},~~
H_I= \sum_{i} c^\dagger_{i\sigma} h^I_{ij,\sigma \sigma^\prime}
c_{j\sigma^\prime}, \nonumber \\
h^K_{ij,\sigma\sigma^\prime}&=& \Big\{-t( \delta_{i,j+\hat x}
+ \delta_{i,j-\hat x} +\delta_{i,j+\hat y}+
\delta_{i,j-\hat y} )-\mu\delta_{ij}\Big\}\delta_{\sigma\sigma^\prime}
\nonumber \\
h^I_{ij,\sigma\sigma^\prime}&=&  U n(i,\tau) \delta_{ij}
\delta_{\sigma\sigma^\prime}. \eea
Here $h^K_{ij,\sigma\sigma^\prime}$
and $h^I_{ij,\sigma\sigma^\prime}$ are defined for both spin
components on each site. After integrating out the fermions, we
obtain \bea\label{B} I+B=I+{\cal T} \exp \big\{-\int^\beta_0 d
\tau~~ (h_K +h_I(\tau)) \big\}. \eea
Note that the matrix kernels $h^K_{ij,\sigma\sigma^\prime}$ and
$h^I_{ij,\sigma\sigma^\prime}$ entering in Eq. (\ref{B}), as well as
the $I+B$ matrix itself, are $2 N\times 2 N$ matrices, if the
lattice system under simulation has $N=L_x\times L_y$ sites. In
the subsequent discussions, we shall simply use the second quantized
operators $H_K$ and $H_I$ interchangeably with the first quantized matrix
kernels $h_K$ and $h_I$ to save some writing, whenever their meanings are
obvious from the context.

In practice, $I+B$ needs to discretized as
\bea\label{IplusB} && I+B =  I +e^{-\Delta\tau H_K} e^{-\Delta\tau
H_{i}(\tau_l)}
e^{-\Delta\tau H_K} e^{-\Delta\tau H_{i}(\tau_{l-1})}...\nonumber\\
&&...e^{-\Delta\tau H_K} e^{-\Delta\tau H_{i}(\tau_{1})},
\eea
where $\Delta \tau=\beta/l$ is the discretized time slice.

Similarly,  at $U>0$,  Eq. (\ref{hubbard12}) can be decomposed in the spin
density channel as
\bea
&&Z=\int D S_z \exp \Big\{ -2 U \int^\beta_0 d\tau
\sum_i S^2_z(i,\tau)
\Big\} \det \{  I+B \}, \nonumber \\
\eea
with the same expression for $B$ as in Eq. (\ref{B}),
but with $H_I$ replaced by
\bea\label{spin-HS}
H_I (\tau)&=& - 2U
\sum_i \Big \{ c^\dagger_{i\alpha}(\tau) \sigma^z_{\alpha\beta}
c_{i\beta}(\tau) \Big \} S_z(i,\tau).
\eea

It is well known that the spin 1/2 Hubbard model is free of the
sign problem either for $U<0$ or for $U>0$ at half-filling and in a
bipartite lattice \cite{HIRSCH1983,HIRSCH1985}. The usual proof is
based on the factorization of the fermion determinant as \bea \det
\{I+B\} =\det \{I +B_{\uparrow}\} \det \{ I+ B_{\downarrow} \}.
\eea In the negative $U$ case, the HS decomposition in Eq.
(\ref{density-HS}) enables such a factorization, and
$B_{\uparrow}$ is identical to $B_{\downarrow}$ for any HS field
configurations. Therefore $\det \{ I+B\}$ is the square of a real
number and thus positive definite. Generally speaking, in the
positive $U$ case, the HS decomposition in Eq. (\ref{spin-HS})
still enables factorization, but $\det \{I +B_{\uparrow}\}$ is
different from $\det \{I +B_{\downarrow}\}$, thus the sign problem
appears. However, at half-filling and on a bipartite lattice, it is
possible to change the sign of $U$ while keeping the kinetic energy
part invariant by  a partial  particle-hole transformation only on
spin down particles \bea\label{partial_ph}
c_{i\uparrow}\rightarrow c_{i\uparrow},~~~~~
c_{i\downarrow}\rightarrow (-)^i c^\dagger_{i\downarrow}, \eea
then the above algorithm is also applicable. Nevertheless, this
transformation can not be applied to lattices which are not
bipartite, or away from the half-filling ($\mu\neq 0$), thus the
sign problem remains in general.

Recently, an anisotropic two band model explicitly breaking the spin
rotational symmetry is also shown to be free of the sign problem
\cite{ASSAAD2003}. The Hamiltonian is defined by
\bea \label{assaad}
H&=&-t \sum_{ij,\sigma} (c^\dagger_{i\sigma} c_{j\sigma}+h.c.) -\mu
\sum_{i,\sigma} n_\sigma(i)
\nonumber \\
&-& |U| \sum_i ( n_1(i)-n_2(i) +n_3(i)-n_4(i) )^2,
\eea
where $n_\sigma(i)=c^\dagger_\sigma (i) c_\sigma(i)$ are particle densities
for each spin components $\sigma=1,2,3,4$.
The interaction part can be decoupled as
\bea
Z&=&\int D S \exp\{- |U| \int_0^\beta  d \tau \sum_i S^2(i,\tau)\} \nonumber \\
&\times&
\exp\{-\int_0^\beta d \tau (H_0 +H_I(\tau)) \} \nonumber \\
H_I(\tau)&=& \sum_{i}
( c^\dagger_{i,1} c_{i,1} -c^\dagger_{i,2} c_{i,2}
+ c^\dagger_{i,3} c_{i,3} -c^\dagger_{i,4} c_{i,4}
) S(i,\tau) \nonumber
\eea
This HS decomposition enables the factorization of the fermion determinant as
\bea
\det \{I+B\}=\det \{I+B\}_{12} \det \{I+B\}_{34},
\eea
where $\det \{I+B\}_{12}$ and  $\det \{I+B\}_{34}$ for spin components $1,2$
and $3,4$ respectively are identical and real. Therefore, the fermion
determinant is positive in this case as well. However, a disadvantage
of this model is the explicit breaking of the spin rotational symmetry.

\section{Fundamental theorem of $T-$invariant decomposition}
\label{sect:theorem}
We now show that the condition of factorizibility of the fermion
determinant is unnecessarily restrictive, and a more general
condition can be precisely stated. The fermion determinant is a
product of all the eigenvalues. Since $I+B$ involves a time
ordered product, it may not be Hermitian and the eigenvalues may
be complex in general. Because the ensemble of HS field
configurations is arbitrary, one would naively not expect any
special relations among the eigenvalues. Surprisingly, the time
reversal symmetry provides an important relationship among the
eigenvalues. To formulate the fundamental theorem,
we consider $H_K$ and $H_I$ in the $I+B$ matrix of Eq.
(\ref{IplusB}) to be the HS decomposed single particle Hamiltonian
matrix derived from a general Hamiltonian, not necessarily the
$s=1/2$ Hubbard model.

{\it Theorem:} If there exists an anti-unitary operator $T$, such that
\bea\label{theorem}
T H_K T^{-1} = H_K,\ \ \ T H_I T^{-1} = H_I, \ \ \ T^2=-1,
\eea
then the eigenvalues of the $I+B$ matrix always appear in complex
conjugate pairs, {\it i.e.}, if $\lambda_i$ is an eigenvalue, then
$\lambda^*_i$ is also an eigenvalue. If $\lambda_i$ is real, it is
two-fold degenerate.  In this case, the
fermion determinant is positive definite,
\bea
\det (I+B) = \prod_i |\lambda_i|^2 \geq 0.
\eea

 Proof: From the condition of the theorem stated in Eq.
(\ref{theorem}), it obviously follows that $T (I+B) T^{-1} =
(I+B)$. For simplicity, we first consider the case where $I+B$ is
an $n\times n$ dimensional diagonalizable matrix, {\it i.e.},
there exists a non-singular matrix $P$ satisfying \bea P^{-1}
(I+B) P=\mbox{diag} \{\lambda_1, \lambda_2, ..., \lambda_n\}. \eea
The $n$ columns of $P$ can be viewed as a set of linearly
independent state-vectors \bea\label{statevectors} P=\{
|\Psi_1\rangle, |\Psi_2\rangle, ..., |\Psi_n\rangle\}, \eea
Suppose that $|\Psi_i\rangle$ is an eigenvector with eigenvalue
$\lambda_i$, {\it i.e.} $(I+B)|\Psi_i\rangle = \lambda_i
|\Psi_i\rangle$. Using the anti-unitary property of $T$, we see
that \bea\label{proof} (I+B) T|\Psi_i\rangle = T(I+B)T^{-1}
T|\Psi_i\rangle = \lambda_i^*T|\Psi_i\rangle. \eea Therefore,
$T|\Psi_i\rangle$ is also an eigenvector, with eigenvalue
$\lambda_i^*$. Since $T^2=-1$, $T|\Psi_i\rangle$ and
$|\Psi_i\rangle$ are orthogonal to each other. This shows that
$\lambda_i$ and $\lambda_i^*$ are two different eigenvalues, thus
the eigenvalues of $I+B$ appear in complex conjugate pairs as
stated in the theorem. If $I+B$ is Hermitian, our theorem reduces
to Kramer's theorem on the time reversal symmetry in quantum
mechanics, stating that the eigenvalues of $I+B$ are real, and
two-fold degenerate.

In the general case, $I+B$ may not be diagonalizable, instead it can
always be transformed into the Jordan normal form as diagonal blocks
\bea \label{jordan}
P^{-1} (I+B) P = \mbox{diag} \{ J_1, J_2,..., J_k \},
\eea
where $P$ is  an $n\times n$ non-singular matrix as before and
$J_i$ is an $l_i\times l_i$ bi-diagonal matrix as
\bea
J_i= \left( \begin{array}{cccccc}
\lambda_i& 1       &   &  & &  \\
         &\cdot& \cdot  &  &    \\
         &     & \cdot&\cdot & \\
         &    &       & \lambda_i& 1 \\
         &    &       &          &\lambda_i
\end{array}
\right). \eea The determinant of $I+B$ is still the product of all
the eigenvalues 
\bea 
\det(I+B) =\prod_{i=1}^{k} (\lambda_i)^{l_i}.
\eea 
As in Eq. \ref{statevectors}, $P$ can be viewed as $n$
linearly independent column state-vectors as 
\bea
P&=&\{ P_1, P_2,...., P_k \}, 
\eea
where each $P_i$ is an $n \times l_i$ matrix
containing $l_i$ column state-vectors 
\bea 
P_i&=&\{|\Psi_{m+1}\rangle, ...,|\Psi_{m+l_i}\rangle\},~~~
m=\sum_{j=1}^{i-1} l_j
\eea
For each Jordan block $J_i$, it satisfies 
\bea 
(I+B) P_i =P_i J_i,
\eea 
thus among the $l_i$ state-vectors in $P_i$, $|\Psi_{m+1}\rangle$
is the only eigenvector with eigenvalule $\lambda_i$.
It is straightforward to show that 
\bea 
(I+B) (T P_i) = (T P_i) J^*_i, 
\eea where $(T P_i)$
is defined as 
\bea 
(T P_i)= \{T|\Psi_{m+1}\rangle,~ ...,~
T|\Psi_{m+l_i}\rangle\}, 
\eea 
and $T|\Psi_{m+1}\rangle$ is the only eigenvector with eigenvalue
$\lambda_i^*$ in $(T P_i)$.
Again since $|\Psi_{m+1}\rangle$ and $T|\Psi_{m+1}\rangle$ are
orthogonal to each other,
$(T P_i)$ contains different
state vectors from what $P_i$ does.
As a result, $J_i$ and $J_i^*$ are different Jordan blocks. 
As before, the Jordan blocks appear in
complex conjugate pairs and so do the eigenvalues. This completes
the proof for the general case of $I+B$.

\begin{figure}
\centering\epsfig{file=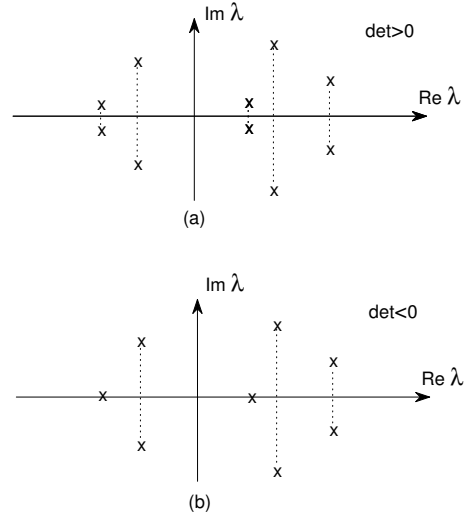,clip=1,width=75mm,angle=0}
\caption {Distribution of eigenvalues in the complex plane. (a)
Eigenvalues of a fermion matrix satisfying the conditions of our
theorem are always paired. (b) Complex eigenvalues of a generic
real matrix are paired, but real eigenvalues are not two-fold
degenerate in general, leading to negative determinants.
}\label{eigenvalues}
\end{figure}

Since the anti-unitary operator $T$ used in our theorem shares
similar properties as the time reversal transformation in quantum
mechanics, we call our method $T$-invariant decomposition. However,
it is important to emphasize that any anti-unitary operator with
the stated mathematical properties could work here. In some
examples we shall discuss,  $T$ does not
have the explicit physical meaning of the time reversal
transformation.

It is also important to point out that the $T^2=-1$ condition is
essential for our theorem. In the case when the fermion matrix is
real, one can define a trivial anti-unitary operator $T=C$,
where $C$ denotes the complex conjugation.
In this case, if the eigenvalue $\lambda_i$ is complex, {\it i.e.},
$\lambda_i\neq \lambda_i^*$, then $\lambda_i^*$ must also be an
eigenvalue. However, when $\lambda_i$ is real, it is in general
not two-fold degenerate, since $|\Psi\rangle$ and $T|\Psi\rangle$
may not be orthogonal for the case of $T^2=1$. In this case, an
odd number of negative eigenvalues would lead to a negative
determinant. The distribution of eigenvalues in the complex plane
for a fermion matrix satisfying the condition of our theorem and
the eigenvalues of a generic real fermion matrix is illustrated in
Fig. \ref{eigenvalues}. When the conditions of our theorem is
violated, either the complex conjugate eigenvalue pairs collide on
the real axis and move off from each other along the real axis, or
the two-fold degenerate eigenvalues move off directly from each
other along the real axis.

A restricted version of our theorem was originally discussed in
the context of nuclear physics\cite{koonin1997}. However, these
authors overlooked the case that $I+B$ may not be diagonalizable,
thus their proof was not complete. In addition, our
$T$-transformation is not restricted to the physical time-reversal
transformation as in Ref.\cite{koonin1997}, thus the theorem
applies to a much wider class of models.

We now illustrate this general theorem for the case of the $s=1/2$
Hubbard model. For the spin 1/2 system on each site, the
time-reversal transformation $T$ is defined as $T(i)=R(i) C$,
satisfying $T^2(i)=-1$ , where
\bea\label{Rmatrix} R= -i \sigma_y
= \left( \begin{array}{cc}
               0  & -1  \\
               1  & 0            \end{array} \right).
\eea
For the entire system,
the time-reversal operator is defined as the direct product
$T= (\Pi_i \otimes R(i)) C$. The four
independent fermion bilinears in the particle-hole channel can be
classified as the particle number $n(i)=\psi^\dagger_{i,\alpha}
\psi_{i,\alpha} $ and spin $\vec
S(i)=\psi^\dagger_{i,\alpha}(\sigma/2)_{\alpha\beta}
\psi_{i,\beta} $, which are even and odd under the $T$
transformation respectively:
\bea \label{eq:timerevs1} T n(i)
T^{-1}= n(i), ~~~~~ T \vec S(i) T^{-1}= -\vec S(i).
\eea
Now we can understand the absence of sign problem in the negative
$U$ case as follows. The density channel decomposition is
$T$-invariant, namely, $T(H_K+H_I(\tau))T^{-1}=H_K+H_I(\tau)$. The
conditions of our theorem is satisfied and the fermion determinant
is thus positive. For $U>0$, the Hamiltonian can be decoupled in
the density channel at the cost of involving the imaginary
number $i$, or decoupled in the spin channel with only real numbers.
In either cases, while $H_K$ is still even under $T$, $H_I$ is odd.
The conditions of our theorem does not apply, and the sign problem
appears in general.

For a general interacting fermion model, we can always express
$T=RC$, where $R R^*=-1$ and $R^*$ is the complex conjugate of $R$.
In many cases, $R$ is purely real, and it reduces to $R^2=-1$. The
general condition for our theorem then reads
\bea\label{canonical} R (H_K+H_I) R^{-1} = (H_K+H_I)^*, \eea with
the unitary matrix $R$ satisfying $R R^*=-1$ for any
configurations of the HS field. Again we emphasis that the precise
form for $R$ in Eq. (\ref{Rmatrix}) is not necessary.

While our new method does not lead to any improvement of the sign
problem for the $s=1/2$ Hubbard model, we shall show now that it
significantly improves the QMC algorithm for multi-band,
multi-layer and higher spin models, since the conditions for our
theorem is far less restrictive than the condition for the
factorizibility of the fermion determinant. Let us illustrate the
general idea here by looking at the example of a two band spin 1/2
model or a spin 3/2 model. In this case, we have fermion operators
$\psi_{i,\beta}$ within one unit cell, where $\beta=1,2,3,4$.
Therefore, there are 16 fermion bilinears, of the form
$M^I=\psi^\dagger_{i,\alpha} M^I_{\alpha\beta} \psi_{i,\beta}$,
where $I=1,...,16$. The 16 $M^I_{\alpha\beta}$ matrices can in general be
expressed in a complete basis in terms of the product of $s=3/2$
matrices $S_i$:
\bea \label{s32_algebra}
& & I, \nonumber \\
& & S^i, \ \ i=1,2,3, \nonumber \\
& & \xi^a_{ij} S_i S_j, \ \ a=1,..,5, \ \ \xi^a_{ij}=\xi^a_{ji}, \ \ \xi^a_{ii}=0, \nonumber \\
& & \xi^L_{ijk} S_i S_j S_k, \ \ L=1,..,7, \ \ \xi^L_{ijk}=\xi^L_{jik}, \ \ \xi^L_{iik}=0,\nonumber \\
\eea
where $\xi$'s are fully symmetric, traceless tensors. If one
insists on the factorizibility of the fermion determinant, one
could only perform the HS decomposition in the density channel
using the identity matrix $I$. However, since the $\xi^a_{ij} S_i S_j$
matrix contains an even power of spin matrices, it is also even under
time reversal. HS decomposition in this channel does not lead to
factorization of the fermion determinant, but according to our general
theorem, it does lead to paired eigenvalues, and therefore, a positive
fermion determinant. As we see from this non-trivial example, our method of
$T$-invariant decomposition is indeed more general and more powerful compared
with the traditional method of factorization. We shall show the enlarged
parameter space for QMC algorithm explicitly in the next section.

\section{ Application in spin 3/2 and $n-1/2$ Hubbard model }
\label{sect:spin32} In this section, we apply the method of
$T$-invariant decomposition to the $s=\frac{3}{2}$ model as an
explicit example, and discuss the sign problem accordingly. After
that, we generalize it to arbitrary fermionic Hubbard models with
$s=n-1/2$. These models are not of only academic interests. In
fact, the rapid progress in ultra-cold atomic systems provides an
opportunity to study higher spin fermions. The simplest cases are
the spin 3/2 atoms, such as $^9$Be, $^{132}$Cs, $^{135}$Ba,
$^{137}$Ba atoms. Another important research direction is the trapped
atoms in an optical lattice, formed by the standing wave laser
beams, where the Hubbard model is a good approximation for these
neutral atoms.

\subsection{The s=3/2 Hubbard model}
The spin 3/2 Hubbard model is defined as\cite{WU2003}
\bea \label{hubbard32}
&&H=-t\sum_{\langle ij\rangle ,\sigma} \big \{ c^\dagger_{i\sigma}
c_{j\sigma} +h.c.\big \}-(\mu+\mu_0)
\sum_{i\sigma} c^\dagger_{i\sigma} c_{i\sigma}  \nonumber \\
&&+U_0 \sum_i P_0^\dagger(i) P_0(i)
+U_2 \sum_{i,m=\pm2,\pm1,0} P_{2m}^\dagger(i) P_{2m}(i), \nonumber  \\
\eea with $\mu_0=(U_0+5 U_2)/4$. $\mu$ is fixed to be zero at
half-filling on a bipartite lattice, to ensure the particle-hole
(p-h) symmetry generated by the transformation
$c_{i,\sigma}\rightarrow (-)^i c^\dagger_{i,\sigma}$. Because of
the Pauli's exclusion principle, only on-site interactions in the
total spin singlet ($S_T=0$) and the quintet ($S_T=2$) channels
are allowed. $P^\dagger_0, P^\dagger_{2m}$ are the singlet and
quintet pairing operators defined by \bea\label{quintetpair}
&&P^\dagger_{0}(i) (P^\dagger_{20}(i))= {1\over \sqrt{2}} \{
c^\dagger_{i,{3\over2}} c^\dagger_{i,-{3\over2}} \mp
c^\dagger_{i,{1\over2}} c^\dagger_{i,-{1\over2}}\},
\nonumber \\
&&P^\dagger_{2,2}(i)= c^\dagger_{i,{3\over2}} c^\dagger_{i,{1\over2}},
\hspace{10mm}
P^\dagger_{2,1}(i)= c^\dagger_{i,{3\over2}} c^\dagger_{i,-{1\over2}},
 \nonumber \\
&&P^\dagger_{2,-1}(i)= c^\dagger_{i,{1\over2}}
c^\dagger_{i,-{3\over2}}, \hspace{1mm} P^\dagger_{2,-2}(i)=
c^\dagger_{i,-{1\over2}} c^\dagger_{i,-{3\over2}}. \eea
The $s=3/2$ Hubbard model has an exact $SO(5)$ or equivalently, $Sp(4)$
symmetry, without any fine tuning of the parameters\cite{WU2003}.
This follows from the fact that singlet and quintet channel
interactions can also be interpreted as $SO(5)$ group's singlet
and 5-vector representations. When $U_0=U_2$, the model has a
larger symmetry, namely the $SU(4)$ symmetry. The $SU(4)$
symmetric Hubbard model has been extensively studied in the
transition metal oxides with double orbital degeneracy
\cite{LI1998}.

To illustrate the $T$-invariant decomposition for this model, we
first define the 4-component spinor
\bea \psi({i})=\big (
c_{3\over2}(i), c_{1\over2}(i), c_{-{1\over2}}(i),
c_{-{3\over2}}(i) \big )^T. \eea
In this representation, we define
five $4\times 4$ Dirac $\Gamma^a$ $(1\le a\le 5)$ matrices to
construct the $Sp(4)$ or $SO(5)$  algebra as \bea \Gamma^1=\left (
\begin{array} {cc}
0 & i I\\
-i I& 0
\end{array} \right) ,
\Gamma^{2,3,4}=\left ( \begin{array}{cc}
{\vec \sigma}& 0\\
0& {-\vec \sigma} \end{array}\right),
\Gamma^5=\left( \begin{array} {cc}
0& -I \\
-I & 0 \end{array} \right ),\nonumber \eea where $I$ and
$\vec{\sigma}$ are the 2$\times$ 2 unit and  Pauli matrices. The
ten $SO(5)$ generators are defined as $\Gamma^{ab}= -{i\over 2} [
\Gamma^a, \Gamma^b] (1\le a,b\le5)$. Since the $SO(5)$ group is
equivalent to the $Sp(4)$ group, there exists a symplectic matrix $R$,
with the properties\cite{SCALAPINO1998}
\bea\label{Rproperty}
&& R^2=-1, \ \ R^\dagger=R^{-1}=~^t R=-R \nonumber \\
&& R \Gamma^a R^{-1}=^t\Gamma^a,\ \ R\Gamma^{ab} R^{-1}=~ -^t\Gamma^{ab}.
\eea
In our explicit representation,
\bea\label{Rspin32}
R=\Gamma^1\Gamma^3= \left (\begin{array}{cc}
0 & -i \sigma_2\\
-i \sigma_2& 0\\
\end{array}\right ).
\eea

Using the $R$ matrix, the $s=3/2$ Hubbard interaction can be written in an
explicitly $SO(5)$ symmetric fashion as
\bea
H&=&-t\sum_{ij} (\psi^\dagger(i) \psi(j) +h.c.)-(\mu+\mu_0) \sum_i
\psi^\dagger(i) \psi(i)\nonumber \\
&+& \frac{U_0}{2} \sum_i \eta^\dagger(i) \eta(i)
+ \frac{U_2}{2}\sum_{i,a} \chi^{\dagger,a}(i) \chi^a(i)
\eea
where $\eta^\dagger(i)=\psi^\dagger(i) (R/2) \psi^\dagger(i)$
is the singlet paring operator, and $\chi^{a,\dagger}(i)=
\psi^\dagger(i) (\Gamma^a R/2) \psi^\dagger(i)$ are the polar forms of the
quintet paring operators in Eq. (\ref{quintetpair}).

In order to implement the method of $T$ invariant decomposition, we first
need to express the interaction terms in the particle-hole channel, rather
than the particle-particle channel. In the particle-hole  channel, there are
$16$  bilinear fermionic
operators, which can be classified into the scalar, vector, and
anti-symmetric tensors (generators) of the $SO(5)$ group as
\bea
n(i)&=& \psi^\dagger_\alpha(i) \psi_\alpha(i)\nonumber \\
n_a(i)&=& \frac{1}{2} \psi^\dagger_\alpha(i) \Gamma^a_{\alpha\beta}
\psi_\beta(i), ~~1\le a \le 5  \nonumber \\
L_{ab}(i)&=&  -\frac{1}{2}\psi^\dagger_\alpha(i)
\Gamma^{ab}_{\alpha\beta} \psi_\beta(i),~~~
1\le a<b\le  5,
\eea
where $n(i)$ is the particle number operator, $n_a(i)$ and $L_{ab}(i)$
represent the spin degrees of freedom.
The ten  $SO(5)$ generators are often conveniently denoted as
\bea\label{so5algebra}
L_{ab}(i)=\left( \begin{array}{ccccc}
0& \Re \pi_x & \Re \pi_y &\Re \pi_z& Q\\
 &   0       & -S_z      & S_y     & \Im \pi_x\\
 &           &  0        & -S_x    & \Im \pi_y \\
 &           &           & 0       & \Im \pi_z \\
 &           &           &         &   0
\end{array} \right).
\eea
Although the similar symbols are used in the $SO(5)$ algebra in the high
T$_c$ cuprates\cite{Demler2004}, operators here have different physical meanings.

These 16 fermion bi-linears  are related through the Fierz identity
\bea\label{fierz1}
\sum_{1\le a<b\le 5} L^2_{ab}(i) + \sum_{1 \le a \le 5}
n_a^2(i) +\frac{5}{4} (n(i)-2)^2=5.
\eea
Defining the time reversal operator as $T=RC$, and using the properties of
the $R$ matrix given in Eq. (\ref{Rproperty}), it can be shown that $n(i)$,
$n_a(i)$ are even while $L_{ab}(i)$ is odd under $T$:
\bea\label{evenodd}
T n T^{-1} =n,~~  T n_a T^{-1} =n_a,~~ T L_{ab} T^{-1}=-L_{ab}.
\eea

On the other hand, we can relate the above $\Gamma$-matrices with the usual
spin $SU(2)$ operators $J_i$, which form a subgroup of the $SO(5)$ group
\bea
J_\pm&=& J_x\pm i~ J_y \nonumber \\
&=&\sqrt 3 (-L_{34}\pm i L_{24})
+ (L_{12}\pm i L_{25}) \mp i(L_{13}\pm i L_{35}),  \nonumber \\
J_z&=&-L_{23} +2~ L_{15} \eea
It is easy to check that the $S_i$ operators form the $s=3/2$
representation of the $SU(2)$ algebra. While the above equation
expresses the spin operators in terms of the $\Gamma$ matrices of
the $SO(5)$ algebra, the reverse can also be accomplished. The
five $\Gamma^a$ matrices can actually be expressed in terms of
quadratic forms of the spin matrices:
\bea
\Gamma^a=\xi^a_{ij} (S_i S_j +S_j S_i)
\eea
where $\xi^a_{ij}$ is a rank-2 symmetric traceless tensor given
Eq. (\ref{s32_algebra}), and discussed more explicitly in
Ref. \cite{murakami2004}.
The ten anti-symmetric tensor $\Gamma^{ab}$ matrices contain both
the three linear (rank-1) $S_i$ and seven cubic symmetric
traceless (rank-3) combination of $S_i S_j S_k$ operators, and they
correspond to the second and the fourth rows of Eq. (\ref{s32_algebra}).
Thus, the $n(i)$ operator describes a particle-hole pair with total
spin zero, the five $n_a(i)$ operators describe five particle-hole
pair states with total spin two, and  the ten $L_{ab}$ operators
include the degenerate three spin-1 and seven spin-3 particle-hole
pair states. From this point of view, the physical meaning of Eq.
(\ref{evenodd}) and Eq. (\ref{s32_algebra}) becomes transparent: 
operators with even total
spins are even under $T$, while operators with odd total spins are
odd under $T$.

Using the identities
\begin{eqnarray}
&&(\Gamma^a R)_{\alpha\beta} (R\Gamma^a)_{\gamma\delta}
= \frac{5}{4} \delta_{\alpha\gamma} \delta_{\beta\delta}
- \frac{3}{4} \Gamma^a_{\alpha\gamma} \Gamma^a_{\beta\delta}
- \frac{1}{4} \Gamma^{ab}_{\alpha\gamma} \Gamma^{ab}_{\beta\delta},
\nonumber\\
&&R_{\alpha\beta} R_{\gamma\delta}
= \frac{1}{4} \delta_{\alpha\gamma} \delta_{\beta\delta}
+ \frac{1}{4} \Gamma^a_{\alpha\gamma} \Gamma^a_{\beta\delta}
- \frac{1}{4} \Gamma^{ab}_{\alpha\gamma} \Gamma^{ab}_{\beta\delta},
\end{eqnarray}
and the Firez identity Eq. (\ref{fierz1}), we can now express the
$s=3/2$ Hubbard model in the following form:
\bea \label{hmlattice2}
H_K&=& -t\sum_i (\psi^\dagger_{i,\sigma} \psi_{j,\sigma}+h.c.)
-\sum_i \mu ~\psi^\dagger_{i\sigma} \psi_{i\sigma}\nonumber \\
H_I&=& -\sum_{i, 1\le a\le 5}\Big \{
\frac{g_c}{2} (n(i)-2)^2+\frac{g_v}{2} n_a^2(i) \Big \}
\eea
where
\bea \label{gc_gv}
g_c &=& -(3 U_0+ 5 U_2)/8  \nonumber \\
g_v &=& (U_2-U_0)/2.
\eea

\subsection{Absence of the sign problem}
After a series of transformations, we arrived at a form of the $s=3/2$ Hubbard
which is suitable for the $T$-invariant decomposition method. The interactions
in Eq. (\ref{hmlattice2}) are fully expressed in $T$-invariant fermion
operators in the particle-hole channel. When $g_c,g_v \ge 0$, {\it i.e},
\bea \label{U0U2}
-3/5~U_0\ge U_2 \ge U_0,
\eea the partition
function can be expressed using the $T$-invariant decomposition as
\bea
&&Z=\int D\psi^\dagger D \psi~ \exp\{ -\int_0^\beta d\tau
~\psi^\dagger_\sigma ( {\partial\over \partial \tau}+H) \psi_\sigma\}
\nonumber \\
&&= \int D n \int D n^a
\exp\Big\{ -\frac{g_c}{2}
\int_0^\beta d\tau \sum_i (n(i,\tau)-2)^2 \nonumber \\
&&-\frac{g_v}{2} \int_0^\beta d\tau \sum_{i,a} n_a^2(i,\tau) \Big \}
\det \Big \{ I+ B \Big \}.
\eea
Again  $I+B= I+ {\cal T} e^{-\int_0^\beta d\tau
(H_K +H_i(\tau))}$ is obtained from the integration of fermion
fields, $n$ and $n_a$ are real HS bose fields.
The time-dependent interaction $H_I(\tau)$  after the HS
transformation is
\bea
H_{I}(\tau)&=& -g_c \sum_i \psi_{i,\sigma}(\tau) \psi_{i,\sigma}(\tau)
n(i,\tau) \nonumber \\
&&-g_v \sum_{i,a} \psi^\dagger_{i,\sigma}(\tau)
\Gamma^a_{\sigma,\sigma^\prime} \psi_{i,\sigma^\prime}(\tau) n_a(i,\tau).
\eea

\begin{figure}
\centering\epsfig{file=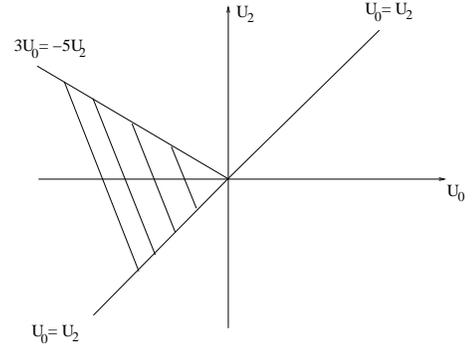,clip=1,width=60mm,angle=0}
\caption {Within the method of $T$-invariant decomposition, the
shaded area marks the parameter region without the sign problem in
the $s=3/2$ Hubbard model at any doping level and lattice
geometry. The fermion determinant can only be factorized along the
$SU(4)$ line with $U_0=U_2<0$, where the traditional algorithms
can be applied without the sign problem. The sign problem along
the line with $U_0=U_2>0$ only disappears at the half-filling and
on a bipartite lattice. }\label{boundary}
\end{figure}

We see that  $H_I(\tau)$ mixes the four spin components together, therefore
the fermion determinant is factorizable if and only if $g_v=0$, which is the
$SU(4)$ line with $U_0=U_2<0$.
We define the time reversal transformation $T$ for the entire
lattice as $T=(\prod_i \otimes R(i)) C$. From Eq. (\ref{evenodd}) we see that
both terms are $T$-invariant:
\bea
T (H_K + H_I) T^{-1} = H_K + H_I,
\eea
and all other conditions of our theorem are met. Therefore, the minus sign
is absent as long as Eq. (\ref{U0U2}) is satisfied. This is a much broader
parameter range shown in Fig. \ref{boundary}, compared to the conventional
factorizibility condition $U_0=U_2<0$.
Our algorithm therefore enables us to study the $s=3/2$ Hubbard model away
from the $SU(4)$ line. Our proof is valid for any filling level
and lattice topology.
In this parameter range, it is shown in Ref. \cite{WU2003} that
a number of interesting competing orders such as the staggered order
of $n^a$, singlet superconductivity, and the charge density wave can
exist there.

At half-filling and on a bipartite lattice where $\mu=0$, the sign
problem also disappears along the $SU(4)$ line at $U=U_0=U_2>0$.
Similar to the spin 1/2 case, after performing a partial
particle-hole transformation
\bea
c_{i\frac{-1}{2}}\rightarrow
(-)^i c^\dagger_{i\frac{-1}{2}}, ~~~~ c_{i\frac{-3}{2}}\rightarrow
(-)^i c^\dagger_{i\frac{-3}{2}},
\eea
while keeping
$c_{i\frac{3}{2}}$, $c_{i\frac{1}{2}}$ operators unchanged. The
kinetic energy part is invariant under the above transformation,
while the interaction part is change into $H_{int}=2 U \sum_i
L_{15}^2 (i)$. It can be decomposed using the imaginary number as
\bea Z&=&\int D Q \exp\Big\{ -2 U\int^\beta_0 d\tau Q^2(i) \Big\}
\det\Big\{ I+B \Big\} \nonumber \eea where $B={\cal T}
e^{-\int^\beta_0 d \tau H_K +H_i(\tau)}$ with $H_{i}(\tau)= i U
\sum_i \psi^\dagger_{i}(\tau) \Gamma^{15} \psi_{i}(\tau)
Q(i,\tau)$. Because $T (i L_{15}) T^{-1}=i L_{15}$, the $\det
(I+B)$ is positive definite. However, we did not succeed to
generalize this at negative values of $g_c$ or $g_v$ away from the
$SU(4)$ line at half-filling.

In practice, it is more efficient to sample with discrete HS
transformation using two Ising-like fields $\eta, s$ for each
quartic fermion term as in Ref. \cite{MOTOME1997} instead of using
the continuous HS boson field. For any  bilinear fermionic
operator $O(i)$, the decomposition below has the numerical
precision at the order of $O(\Delta \tau)^4$ as \bea &&e^{g \Delta
\tau \hat O(i,\tau)^2}=\sum_{l,s=\pm1} {\gamma_l\over4}
e^{s\eta_l\sqrt{\Delta\tau g}
\hat O(i,\tau)}+O(\Delta\tau^4),  \nonumber \\
&&e^{-g \Delta \tau
\hat O(i,\tau)^2}=\sum_{l,s=\pm1} {\gamma_l\over4}
e^{i s\eta_l\sqrt{\Delta\tau g} \hat O(i,\tau)}+O(\Delta\tau^4),
\nonumber, \eea
where $g>0$ and  $\gamma_l=1+{\sqrt{6}\over 3} l,
\eta_l=\sqrt{2(3-\sqrt{6}l)}$.
The above proof for the positive definite of $\det(I+B)$ applies
equally well in this scheme.

We only used the time reversal properties of the $SO(5)$ algebra in
the above proof; the exact $SO(5)$ symmetry is useful for
transforming the model expressed in the particle-particle channel
to the particle-hole channel, but is not essential. A general
anisotropic spin 3/2 lattice model is defined by
\bea\label{generalmodel}
H&=&-\sum_{\avg{ij,a }} \big\{ t~\psi^\dagger_{i\alpha}
\psi_{j\alpha} +t_a~\psi^\dagger_{i\alpha} \Gamma^a_{\alpha\beta}
\psi_{j\beta}
+h.c. \big \} \nonumber \\
&+& \sum_{i,a}  \big\{ h_a  n_a(i)-\mu~ n(i)\big \}
+\sum_{i,a<b} \Big \{
-\frac{g_c}{2} (n(i)-2)^2 \nonumber \\
&-&\frac{g_a}{2} n_a^2(i) + \frac{g_{ab}}{2} L_{ab}^2(i) \Big \},
\eea
where $t_a$ is the spin dependent hopping amplitude,
$h_a$ is the analogy of the Zeeman field coupling to the
$n_a(i)$ field, $g_a$ and $g_{ab}$ are coupling constants in
corresponding channels. When $g_c, g_a, g_{ab}$ are arbitrary
positive interaction parameters, we can perform the same
decomposition process as before. By using the fact that $T (i
L_{ab}) T^{-1}= i L_{ab}$, we again reach the positive definite
fermion determinant. This conclusion also holds for any valid
representation of $\Gamma$ matrices, with the redefined $n(i),
n_a(i), L_{ab}(i)$ and time reversal operations accordingly.

\subsection{General higher spin Hubbard models}
We can generalize the results in the spin 3/2 case to the
any fermionic system with spin $s=n-\frac{1}{2}$.
The spin $s=n-\frac{1}{2}$ Hubbard model can be written as
\bea \label{spin2n}
&&H=-t\sum_{\langle ij\rangle ,\sigma} \big \{ c^\dagger_{i\sigma}
c_{j\sigma} +h.c.\big \}-(\mu+\mu_0)
\sum_{i\sigma} c^\dagger_{i\sigma} c_{i\sigma}  \nonumber \\
&&+ \sum_{i,J,J_z} U_J P^\dagger_{J J_z}(i)  P_{J J_z}(i), \eea
where $J=0, 2, ..., 2n-2$ are the total spin of the
particle-particle pairs, $J_z=0,\pm 1, ..., \pm J$. The pairing
operators $P^\dagger_{J,J_z}$ are defined through the
Clebsch-Gordan coefficient for two indistinguishable particles as
\bea P_{J,J_z}^\dagger(i) = \sum_{\alpha\beta}
\avg{J,J_z|s,s;\alpha\beta} c^\dagger_\alpha(i) c^\dagger_\beta(i)
\eea The total spin of the particle-particle pair takes only even
integer values so that Pauli principle is satisfied on every site.
At half-filling and on a bipartite lattice, $\mu=0$ ensures the
particle-hole symmetry, and $\mu_0= 1/(2n) \sum_J (2J+1) U_J$.

The general strategy to implement the method of $T$-invariant
decomposition is to first transform the interaction terms
originally expressed in the particle-particle channel to the
particle-hole channel. In this case, we have fermion operators
$\psi_{i,\beta}$ within one unit cell, where $\beta=1,..,(2s+1)$.
Therefore, there are $(2s+1)^2$ fermion bi-linears, of the form
$M^I=\psi^\dagger_{i,\alpha} M^I_{\alpha\beta} \psi_{i,\beta}$,
where $I=1,...,(2s+1)^2$. The $(2s+1)^2$ $M^I_{\alpha\beta}$ matrices can
in general be expressed in a complete basis in terms of the
product of spin $s$ matrices $S_i$:
\bea \label{s_algebra}
& & 1, \nonumber \\
& & S^i, \ \ i=1,2,3, \nonumber \\
& & \xi^a_{ij} S_i S_j, \ \ a=1,..,5, \nonumber \\
& & ... \nonumber \\
& & \xi^L_{i_1,i_2,... i_J} S_{i_1} S_{i_2} \cdot \cdot \cdot
S_{i_J}, \ \ L=1,..,(4s+1), \nonumber \\
\eea where $\xi$'s are fully symmetric, traceless tensors,
satisfying \bea \label{tensor1} \xi^L_{i_1,i_2,...
i_J}=\xi^L_{i_2,i_1,... i_J} \eea or any other permutation of
indices, and \bea \label{tensor2} \xi^L_{i_1,i_1,... i_J}=0 \eea
Spherical harmonics can be used to explicitly construct these
tensors\cite{HO1999}. This decomposition is obviously complete,
since \bea \label{complete} (2s+1)^2=1+3+5+...+(4s+1) \eea
According to the method of $T$-invariant decomposition, any negative
interaction terms in the even spin channel like $1$, $\xi^a_{ij}
S_i S_j$, ... or any positive interaction terms in the odd spin
channel like $S_i$, $\xi^L_{ijk} S_i S_j S_k$, ... can be
simulated by our algorithm without the sign problem.

In the following, we shall illustrate this general procedure more
explicitly for a special case of the higher spin Hubbard model
where \bea \label{sp2n} U_2=U_4=...=U_{2n-2}\equiv U'. \eea The
generic higher spin Hubbard model only has the spin $SU(2)$ symmetry
for $s\neq\frac{3}{2}$. However, under the above condition, the higher
spin Hubbard has the $Sp(2n)$ symmetry. When an additional
condition, namely $U_0=U'$ is imposed, the model has a larger,
$SU(2n)$ symmetry. In Appendix \ref{spgroup}, an introduction to the
$Sp(2n)$ algebra is given. As shown there, the singlet pairing
operator is also the singlet of the $Sp(2n)$ group, while all
other $2n^2-n-1$ pairing operators with $J=2,4, ..., 2n-2$
together form a representation for the $Sp(2n)$ group. Thus we
conclude that Eq. (\ref{spin2n}) is $Sp(2n)$ symmetric if and only if
coupling constants satisfy Eq. (\ref{sp2n}). For n=1 and 2, the
$Sp(2n)$ symmetry is generic and does not need any fine-tuning.
Actually, $Sp(2)$ is isomorphic to $SU(2)$, while $Sp(4)$ is
isomorphic to $SO(5)$.  This is consistent with our earlier
finding that the $s=3/2$ Hubbard model has the $SO(5)$, or the
$Sp(4)$ symmetry without any conditions on the parameters\cite{WU2003}.

To show the $Sp(2n)$ symmetry explicitly, we can rewrite the Hamiltonian
in Eq, (\ref{spin2n}) as
\bea
H&=&-t\sum_{\langle ij\rangle ,\sigma} \big \{
\psi^\dagger_{i,\alpha} \psi_{j,\alpha}+h.c.\big \}-\mu
\sum_{i}\psi^\dagger_{i,\alpha} \psi_{i,\alpha}  \nonumber \\
&+& c_0 \sum_i(\psi^\dagger_{i,\alpha} \psi_{i,\alpha}-n)^2
+ c_2\sum_i (\psi^\dagger_{i,\alpha} Y^a_{\alpha\beta} \psi_{i,\beta})^2,
\nonumber \\
c_0&=&\frac{n+1}{4n^2} U_0 + \frac{2n^2-n-1}{4n^2} U^\prime, ~~~
c_2=\frac{U_0-U^\prime}{2n}.
\eea
The expression for $Y^a(1\le a \le 2n^2-n-1)$ is given in
Appendix \ref{spgroup}, where it is also shown that they are even
under the time-reversal transformation.
By the same reasoning before, we perform the HS decoupling in the
above two channels.
Then the sign problem is absent at both $c_1$ and $c_2$ are negative,
{\it i.e.}
\bea
U_0\le U^\prime \le -\frac{n+1}{2n^2-n-1} U_0.
\eea
Because the $Y^a~(1\le a \le 2n^2-n-1)$ are even under time-reversal
transformation while the spin operators $S_i (i=x,y,z)$ are odd,
$Y^a$ can be expanded in the basis of Eq. (\ref{s_algebra}) including
all terms with even powers of spin matrix.

\section{Bi-layer s=1/2 models}\label{sect:SZH}
\begin{figure}
\centering\epsfig{file=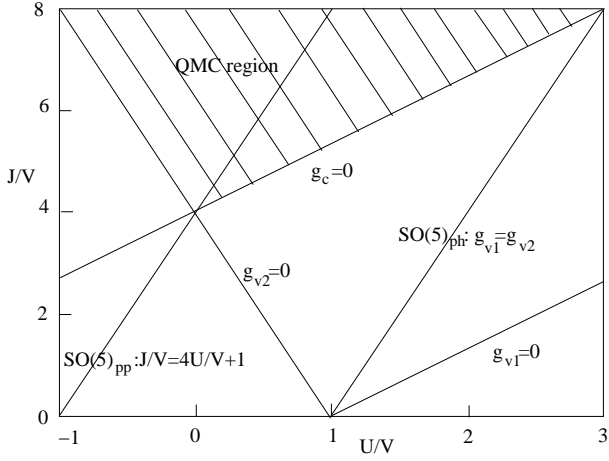,clip=1,width=80mm,angle=0}
\caption{Two $SO(5)$ lines are shown in the SZH model as well as
the QMC region without the sign problem for any filling (hatched
area): $g_{v1}>0, g_{v2}>0$ and $g_c>0$. There is another region
with $V<0$ (not shown). }\label{fig:szhbound}
\end{figure}

The spin 3/2 Hubbard model has a close relationship with a bi-layer model
introduced by Scalapino, Zhang and Hanke(SZH)\cite{SCALAPINO1998}. This
model was constructed and extensively investigated because of the exact
particle-particle channel $SO(5)$ symmetry  between the
antiferromagnetism and superconductivity when the coupling
constants satisfy a simple
relation\cite{BOUWKNEGT1999,DUFFY1998,LIN1998,FRAHM2001}. The
original SZH model was introduced on a two leg ladder, and it
is straightforward to generalize it to a bi-layer system as
\begin{widetext}
\bea\label{SZHham}
H&=&-t_\pp \sum_{\avg{ij}} \big\{c^\dagger_{i\sigma} c_{j\sigma}
+d^\dagger_{i\sigma} d_{j,\sigma}+h.c.\big\}-t_\perp \sum_i
\big \{c^\dagger_{i,\sigma} d_{i,\sigma} + h.c. \big \}
-\mu\sum_i \big\{ n_c (i)+n_d(i) \big \}\nonumber \\
&+& U\sum_{i} \Big\{ (n_{\uparrow,c}(i)-\frac{1}{2})
(n_{\downarrow,c}(i)-\frac{1}{2})
+ (n_{\uparrow,d}(i)-\frac{1}{2})
(n_{\downarrow,d}(i)-\frac{1}{2}) \Big \}\nonumber\\
&+&V\sum_{i}  (n_{c}(i)-1)(n_{d}(i)-1)
+J\sum_{i}  \vec{S}_{i,c}\cdot \vec{S}_{i,d},
\eea
\end{widetext}
where $c^\dagger$ and $d^\dagger$ are creation operators in the upper
and lower layer respectively, $t_\pp$ and $t_\perp$ are the hopping
amplitude in the layer and cross the rung respectively,
$U$ is the onsite interaction, $V$ and $J$ are the charge and Heisenberg
exchange interaction across the rung respectively.
The SZH model is known to have an exact $SO(5)$ symmetry when
\bea\label{so5-pp}
J=4(U+V), \ \ \ \mu=0,
\eea
which unifies antiferromagnetism with superconductivity \cite{SCALAPINO1998}.
Remarkably, there exists another exact $SO(5)$ symmetry in the particle-hole
channel  when
\bea\label{so5-ph}
J=4(U-V), \ \ t_\perp=0,
\eea
and the symmetry is valid for all filling factors.
We denote the former particle-particle $SO(5)$ symmetry as $SO(5)_{pp}$
and the later p-h $SO(5)$ symmetry as $SO(5)_{ph}$.
The two $SO(5)$ symmetric lines are shown in Fig. \ref{fig:szhbound}.
In order to employ the method of $T$-invariant decomposition,
we adopt the view from the $SO(5)$ symmetry in the particle-hole channel
in this section.

\begin{figure}
\centering\epsfig{file=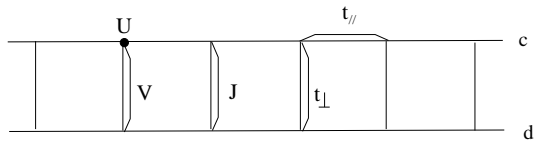,clip=1,width=70mm,angle=0}
\caption{The SZH model defined on a two-leg ladder segment of the
double-layer spin 1/2 system. }\label{SZHmod}
\end{figure}

There are four single fermion states per unit cell in both the
$s=3/2$ Hubbard model and the $s=1/2$ bi-layer model, a mapping between
them can be established through
$\psi_i=(c_{i,\frac{3}{2}},c_{i,\frac{1}{2}},
c_{i,-\frac{1}{2}}, c_{i,-\frac{3}{2}})^T \leftrightarrow
(c_{i,\uparrow}, c_{i,\downarrow}, d_{i,\uparrow}, d_{i,\downarrow})^T$.
We denote the time reversal operator defined
in Eq. (\ref{Rspin32}) for the $s=\frac{3}{2}$ system as $T_1$,
and the usual definition for $s=1/2$ system as $T_2$.
$T_1$ actually is the combined operation of $T_2$ and the interchange
between the upper and lower layers:
\bea
T_1= \left(\begin{array}{cc}
0&I\\
I&0\\
\end{array}\right)
T_2
\eea

The 16 p-h channel fermionic bilinear forms are mapped onto
\bea
n(i)&=& c^\dagger_{i\sigma} c_{i\sigma}+d^\dagger_{i\sigma} d_{i\sigma}
\nonumber \\
n_1(i)& =&-i (d^\dagger_{i\sigma}  c_{i\sigma}-h.c.)/2\nonumber \\
n_5(i) &=&   (d^\dagger_{i\sigma}  c_{i\sigma}+h.c)/2 \nonumber \\
n_{2,3,4}(i) &=& c^\dagger_{i\alpha} (\frac{\vec \sigma}{2})_{\alpha\beta}
c_{i\beta} - d^\dagger_{i\alpha} (\frac{\vec \sigma}{2})_{\alpha\beta}
d_{i\beta}\nonumber \\
\Re \vec \pi (i)&=&  c^\dagger_{i\alpha} (\frac{\vec\sigma}{2})_{\alpha\beta}
d_{i\beta}+h.c. \nonumber \\
\Im\vec\pi (i)&=& -i (c^\dagger_{i\alpha} (\frac{\vec \sigma}{2})_{\alpha\beta}
d_{i\beta}-h.c.) \nonumber \\
\vec S_{i}&=& c^\dagger_{i\alpha}(\frac{\vec \sigma}{2})_{\alpha\beta}
c_{i\beta} +d^\dagger_{i\alpha}(\frac{\vec \sigma}{2})_{\alpha\beta}d_{i\beta}
\nonumber\\
Q&=&(n_c(i)-n_d(i))/2,
\eea
where $n_{1,5}$ are the singlet rung current and rung bond order parameters
respectively, and $\Im \vec \pi$ and $\Re \vec \pi$ are their triplet
counterpart;
$n_{2,3,4}$ are the rung Neel order parameter, and $\vec S$ is the total
rung spin, $Q$ is charge density wave order parameter.
$n$, $n_{1\sim 5}$ are even under the $T_1$ transformation
and the others are odd.
In contrast, $n$, $n_5$, $\Im \vec \pi$ and $Q$ are even under
the usual definition $T_2$ and the others are odd.

The general SZH model can be mapped into an anisotropic $SO(5)$ model
in the following form:
\bea\label{szh_so5}
H&=&-t_\pp \sum_{\langle i j \rangle} \psi_{i\alpha}^\dagger \psi_{j\alpha}
+t_\perp\sum_i \psi_{i\alpha}^\dagger \Gamma^5_{\alpha\beta} \psi_{i\beta}
-\mu \sum_i n_i, \nonumber\\
&+&\sum_{i}\Big\{ -\frac{g_c}{2} (n(i)-2)^2
-\frac{g_{v1}}{2} (n^2_1(i)+n^2_5(i))
-\frac{g_{v2}}{2} \nonumber \\
&\times& (n^2_2(i) +n^2_3(i)+ n^2_4(i) ),
\eea
with
\bea
4 g_c & = & \frac{3}{4} J -U-3V, \ \ \
4 g_{v1} =  \frac{3}{4}J-U+V, \nonumber \\
4 g_{v2} & = & \frac{J}{4}+U-V.
\eea
The particle-hole channel $SO(5)_{ph}$ symmetry is restored at $g_{v1}=g_{v2}$
and $t_\perp=0$, {\it i.e.}, when the conditions of Eq. (\ref{so5-ph})
are satisfied.
At this point, the SZH model expressed in Eq. (\ref{szh_so5}) takes
exactly the same form as the $s=3/2$ Hubbard model expressed in
Eq. (\ref{hmlattice2}).
The equivalence between the two models is therefore rigorously established.

For the general SZH model, the interactions can be expressed purely in
terms of the fermion bi-linears which are invariant under $T_1$ transformation
from  Eq. (\ref{szh_so5}), by virtue of Eq. (\ref{evenodd}).
We perform the $T_1$ invariant decomposition of the interactions
in the region of $g_c, g_{v1},g_{v2} \ge0$,
{\it i.e.}
\bea\label{eq:SZHbound}
\left\{ \begin{array}{l}
\frac{3}{4} J+V\ge U\ge -\frac{1}{4} J +V \\
\frac{3}{4} J\ge U+3 V
\end{array}\right. ,
\eea
as shown in Fig. \ref{fig:szhbound}, then the sign problem is absent.
In this region with positive $g_{v1,2}$ and $g_c$, we expect the
competing orders of the five-vector channel and the superconductivity,
{\it i.e.} the antiferromagnetism, staggered current and the rung-singlet
superconductivity, which can be investigated systematically with high
numerical accuracy, at and away from the half-filling.

\begin{figure}
\centering\epsfig{file=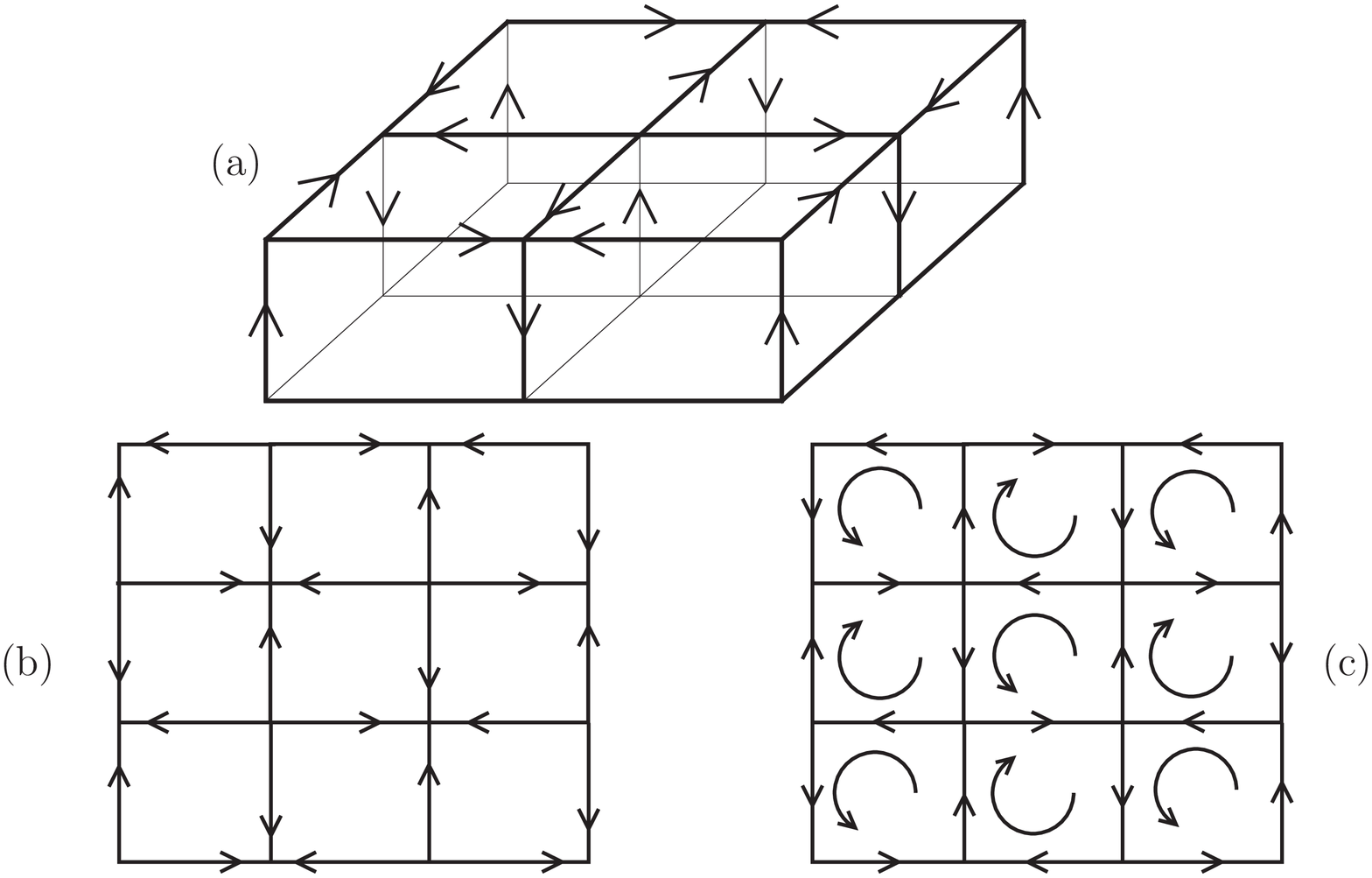,clip=1,width=80mm,angle=0}
\caption{(a) Sketch of a staggered interlayer current phase from
Ref. \cite{CAPPONI2004}. For clarity, we do not show the bottom
layer current. (b)Top view of the bi-layer current. (c) Sketch of
the D-density wave current pattern for comparison. }
\label{fig:current}
\end{figure}

The above algorithm has been applied to demonstrate the
existence of 2-dimensional staggered current phase conclusively
at half-filling with $t_\pp=1, t_\perp=0.1, U=0, V=0.5, J=2$
\cite{CAPPONI2004}.
The current pattern is illustrated in Fig. \ref{fig:current} with
staggered inter-layer currents (SIC) between the bi-layers and
alternating source to drain currents within the bi-layers.
Viewed from the top, this current pattern has a $s$-wave symmetry.
While the D-density wave \cite{CHAKRAVARTY2001A}
currents are divergence free within the layer,
the SIC current is curl free within the layer.
These two  patterns can be considered as dual to each other in two dimensions.
As far as in our knowledge, this is the first time a current
carrying ground state has been conclusively demonstrated in a 2D system.

The mapping between the SZH model and the $s=3/2$ model are not unique.
More generally the SZH model can be written as
\bea\label{recipe}
H&=&-t_\pp \sum_{\langle i j \rangle} \psi_{i\alpha}^\dagger \psi_{j\alpha}
+t_\perp\sum_i \psi_{i\alpha}^\dagger \Gamma^5_{\alpha\beta} \psi_{i\beta}
-\mu \sum_i n_i, \nonumber\\
&+&\sum_{i}\Big\{ -\frac{g_c}{2} (n(i)-2)^2
-\frac{g_{v1}}{2} (n^2_1(i)+n^2_5(i))
-\frac{g_{v2}}{2} \nonumber \\
&\times& (n^2_2(i) +n^2_3(i)+ n^2_4(i) )+
\frac{g_{t1}}{2} Q^2(i)+\frac{g_{t2}}{2} (\vec S(i)\cdot \vec S(i)) \nonumber\\
&+& \frac{g_{t3}}{2} ( \Re \vec \pi(i)\cdot \Re \vec \pi(i)+
\Im \vec \pi(i)\cdot \Im \vec \pi(i)).
\eea
Only three out of the six coupling constants are independent,
as shown here in the correspondence to the $U,V,J$ parameters
\bea
U&=& -4g_c +3  g_{v2}+ g_{t1}- 3 g_{t2} \nonumber \\
V&=& -4g_c +g_{v1}-g_{t1}-3 g_{t3} \nonumber \\
J&=&4(g_{v1}+g_{v2}+g_{t2}+g_{t3}).
\eea
If the $g_{t1}, g_{t2},g_{t3}$ are set to zero, it returns to
Eq. (\ref{szh_so5}).
For any given values for $U,V,J$, if we can find a set of
value of $g_c, g_{v1},g_{v2}, g_{t1}, g_{t2},g_{t3} \ge0$,
then we can perform the HS transformation keeping the invariance under
the $T_1$ operation and arrive at the absence of the sign problem
regardless the doping and lattice topology.
This general decoupling scheme extends the valid parameter region
in Eq. (\ref{eq:SZHbound}).
On the other hand, we can also consider to perform HS decoupling with
the invariance under the usual definition of the time reversal
operation $T_2$.
After setting  $g_{v1}, g_{t3}=0$, we have the condition that
$g_c\ge 0, g_{v2}\le 0, g_{t1}\le 0, g_{t2}\ge 0$.
This decoupling scheme based $T_1$ also enlarges the region of
Eq. (\ref{eq:SZHbound}).
For example, the usual bilayer negative $U$ Hubbard model with $U<0, V=J=0$
is out of that region.
Nevertheless, we can still show the absence of the sign problem
by setting $g_c=-g_{t1}/4=-U/2>0$ and all other parameters zero.

In contrast, the conventional algorithm based on the factorization
of the fermion determinant works only at either $g_{v1}=g_{v2}=0, g_c<0$
or the usual negative U Hubbard model $U<0, V=J=0$. This parameter set
is included in the above HS decomposition schemes respecting either the
$T_2$ or the $T_1$ time reversal symmetry.
We therefore see the significant improvement provided by the method
of the $T$-invariant decomposition.

\section{Models with bond interactions} \label{sect:exotic}
So far, the models we considered only have on-site interactions.
In this section, we will generalize them to include interactions
defined on the bond. Such models can have many exotic phases.

We first consider the following general single layer spin $1/2$  Hamiltonian
with bond interactions
\bea\label{Ham2}
H&=& -t\sum_{\avg{ij},\sigma}(c^\dagger_{i\sigma} c_{j\sigma} +h.c.)
-\mu\sum_i n(i)\nonumber \\
&+& \sum_{\avg{ij}}
\Big\{ - \frac{g_{sbd}}{2}  M_{ij} M_{ij}
+\frac{g_{scur}}{2} N_{ij} N_{ij} \nonumber \\
&+& \frac{g_{tbd}}{2}  \vec M_{ij} \cdot \vec  M_{ij}
-\frac{g_{tcur}}{2}
\vec N_{ij} \cdot \vec N_{ij} \Big \}, \nonumber \\
\vec M_{ij}&=& c^\dagger_{i\alpha}(\frac{\vec \sigma}{2})_{\alpha\beta}
c_{j\beta}+h.c. ,~~
\vec N_{ij}= i\{ c^\dagger_{i\alpha}(\frac{\vec \sigma}{2})_{\alpha\beta}
c_{j\beta}-h.c.\}, \nonumber \\
M_{ij}&=& c^\dagger_{i\sigma} c_{j\sigma}+h.c. ,~~~
N_{ij}= i\{c^\dagger_{i\sigma} c_{j\sigma}-h.c.\},
\eea
where $M_{ij}$ and $N_{ij}$ are the singlet bond and current operators
on the bond $ \avg{ij}$, $\vec M_{ij}$ and $ \vec N_{ij}$ are their
triplet counterparts,
$g_{sbd}$, $g_{scur}$, $g_{tbd}$ and $g_{tcur}$ are the coupling
constants in the corresponding channels.
The sites $i$ and $j$ forming the bond $\avg{ij}$ are not necessary
nearest neighbors, but can be at arbitrary distance apart.
Under the time reversal transformation $T$, $M_{ij}$ and $\vec N_{ij}$
are even while $\vec M_{ij}$ and $N_{ij}$ are odd.

These four interactions are not independent, and can be reorganized
into
\bea
H_{int}&=& \sum_{\avg{ij}}\Big\{  -J_c (c^\dagger_{i\uparrow}
c^\dagger_{i\downarrow}
c_{j\downarrow} c_{j\uparrow} +h.c.)  \nonumber \\
&+& V  (n(i)-1) (n(j)-1)
+J_s \vec S(i) \cdot \vec S(j), \Big\}\nonumber \\
J_c&=& 2(g_{sbd}+g_{scur})+3(g_{tbd}+g_{tcur} ),\nonumber \\
V&=& \frac{g_{sbd}-g_{scur}}{2}-\frac{3}{4} (g_{tbd}-g_{tcur}),  \nonumber \\
J_s&=& 2(g_{sbd}-g_{scur}) + (g_{tbd}-g_{tcur}).
\eea
The $J_c$ term is the pair hopping, $V$ is the charge interaction between
site $i$ and $j$, and $J_s$ is the Heisenberg exchange.
When all of  $g_{sbd}, g_{scur}, g_{tbd}, g_{tcur}$ are positive,
we perform the HS decomposition in each channel respectively as
\begin{widetext}
\bea
Z&=&\int D M  D \vec M D N  D \vec N
\exp \Big\{- \int^\beta_0 d\tau
\sum_{\avg{ij}} g_{sbd}  M^2_{ij}(\tau)
+ g_{scur}  N^2_{ij}(\tau)
+ g_{tbd}  \vec M^2_{ij}(\tau) + g_{tcur} \vec  N^2_{ij}(\tau)^2
\Big\} \nonumber \\
&\times& \det \{  I+B \},~~~~~
\eea
where $I+B= I+{\cal T} e^{-\int^\beta_0 d \tau H_K +H_I(\tau)}$.
$H_I(\tau)$ after the HS decoupling is given by
\bea
H_{I}(\tau)&=& -\sum_{\avg{ij}}  g_{sbd} M_{ij}(\tau)
( c^\dagger_{i,\sigma} c_{j,\sigma} +h.c. )
+ i \sum_{\avg{ij}} g_{scur} N_{ij} (\tau)
i( c^\dagger_{i,\sigma} c_{j,\sigma} -h.c. )\nonumber \\
&-& i \sum_{\avg{ij}}  g_{tbd} \vec M_{ij}(\tau)
( c^\dagger_{i\alpha}(\frac{\vec \sigma}{2})_{\alpha\beta} c_{j\beta} +h.c. )
+ \sum_{\avg{ij}} g_{tcur} \vec N_{ij} (\tau)
i( c^\dagger_{i,\alpha}(\frac{\vec\sigma}{2})_{\alpha\beta} c_{j,\beta} -h.c. )
\eea
\end{widetext}
Therefore, $H_{I}$ and $I+B$ are even under the time-reversal
transformation and  the sign problem is absent.

\begin{figure}
\centering\epsfig{file=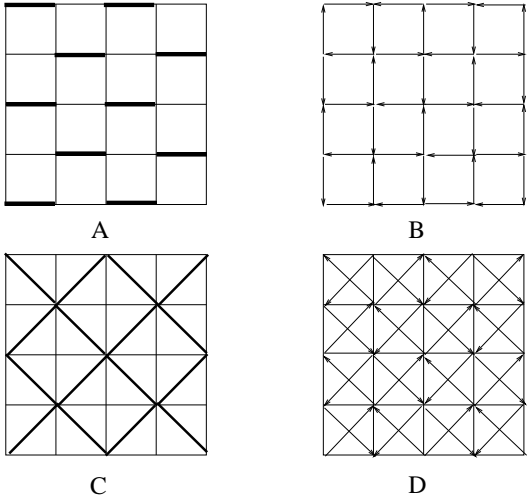,clip=1,width=70mm,angle=0}
\caption{Four possible density-wave phases can be simulated without
sign problem. A) singlet spin-Peierls (p-density wave), B) Triplet
$d_{x^2-y^2}$ density wave, C) singlet $d_{xy}$ density-wave, D)
Triplet diagonal current. }\label{orders}
\end{figure}

The valid parameter region for the above algorithm is very general
as long as  all  $g_{sbd}, g_{scur}, g_{tbd}, g_{tcur}\ge 0$.
As a result, $V$ and $J$ can be either positive or
negative while $J_c$ has to be positive. Many interesting
competing orders are supported in this parameter region. For
example, various density-wave states exist on a square lattice
near half-filling \cite{NAYAK2000} as shown in Fig.
\ref{orders}. With $g_{sbd}, g_{tcur}>0$, the above algorithm
provide a good opportunity to study the singlet bond and the
triplet current order parameters formed by $M_{ij}$ and $\vec
N_{ij}$, while its not  good for study the singlet current and the
triplet bond  order parameters formed by $N_{ij}$ and $\vec
M_{ij}$ because $g_{tbd}, g_{scur}>0$. After setting
$g_{tbd}=g_{scur}=0$, for the bond $\avg{ij}$ connecting the
nearest sites,  the $g_{sbd}$ term favors the $p$-density wave
(spin-Peierls) phase, and the $g_{tcur}$ term favors the triplet
channel $d_{x^2-y^2}$-density wave. The latter order is recently
proposed as the origin as the pseudogap in the high $T_c$
cuprates\cite{LIU2004}. For the bond interaction between the next
nearest bond, {\it i.e.} the diagonal bond, the $g_{sbd}$ term
leads to the singlet $d_{xy}$ order, and  $g_{tcur}$ term leads to
the triplet diagonal current order. The triplet diagonal current
phase was studied in the two-leg ladder system using the
bosonization method in Ref. \cite{wu2003a} and also under the name
of the triplet $F$-density wave in Ref. \cite{TSUCHIIZU2002}.

\begin{figure}
\centering\epsfig{file=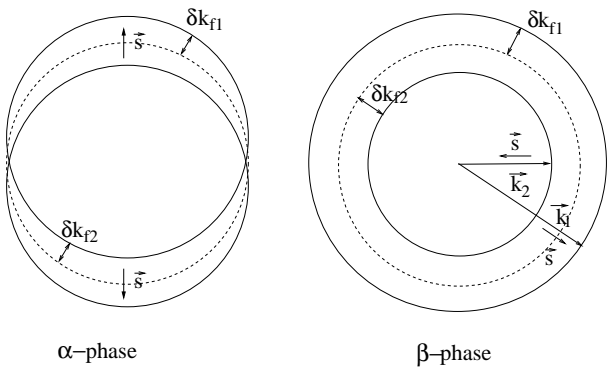,clip=1,width=80mm,angle=0}
\caption{The fermi surface instability in the $F^a_1$ channel,
with dashed lines marking the fermi surface before symmetry
breaking. In the $\alpha$-phase, the anisotropic fermi surface
distortion appears for two spin components. In the $\beta$-phase,
spin-orbital coupling is generated dynamically and two fermi
surfaces are characterized by helicity. }\label{spinorbit}
\end{figure}

When the Fermi surface nesting effect is not important either at
large doping or in the non-bipartite lattice, the $g_{tcur}$ term
can lead to the $F^a_1$  channel of the Landau-Pomeranchuk
instability on the Fermi surface, which was studied recently in
the continuum model in Ref. \cite{wu2004}. After the symmetry
breaking, two possible phases are named as $\alpha$ and $\beta$
phases in analogy to the $A$ and $B$ phases in the triplet
p-wave channel superfluid phase in $^3$He as shown in Fig.
\ref{spinorbit}. The $\alpha$-phase was studied by Hirsch
\cite{HIRSCH1990,HIRSCH1990a} under the name of spin-split phase
on the lattice system with an opposite anisotropic Fermi surface
distortions for two spin components. In contrast, the fermi
surface distortion is isotropic and a spin-orbit coupling is
dynamically generated in the $\beta$ phase. The two single
particle bands are characterized by the helicities. It would be
interesting to study these exotic phases in our version of the
QMC algorithm free of the sign problems.

Bond interactions can also be added into the spin 3/2 Hubbard model of
Eq. (\ref{hubbard32}) as
\bea
H_{bond}&=& \sum_{\avg{ij}}
\Big\{ - \frac{g_{sbd}}{2}  M_{ij} M_{ij}
+\frac{g_{scur}}{2} N_{ij} N_{ij} \nonumber \\
&+& \sum_a -\frac{g_{vbd}}{2}  M^a_{ij} M^a_{ij}
+\frac{g_{vsur}}{2} N_{ij} N_{ij} \nonumber \\
&+& \sum_{a<b} \frac{g_{tbd}}{2}
M^{ab}_{ij}  M^{ab} _{ij}  -\frac{g_{tcur}}{2}
N^{ab}_{ij}  N^{ab} _{ij} \Big \}, \\
M_{ij}&=& \psi^\dagger_i \psi_j+h.c. ,~~~
N_{ij}= i\{\psi^\dagger_i \psi_j-h.c.\} ,\nonumber \\
M^a_{ij}&=& \psi^\dagger_i \frac{\Gamma^a}{2} \psi_j+h.c. ,~~~~
N^a_{ij}= i\{ \psi^\dagger_i \frac{\Gamma^a}{2}\psi_j-h.c.\}, \nonumber \\
M^{ab}_{ij}&=& \psi^\dagger_i \frac{\Gamma^{ab}}{2}\psi_j+h.c. , ~~~
N^{ab}_{ij}= i\{ \psi^\dagger_i \frac{\Gamma^{ab}}{2}
\psi_j-h.c.\}, \nonumber \\
\eea
where $M_{ij}$ and $N_{ij}$ are the singlet bond and current operators
on the bond $ \avg{ij}$, $ M^a_{ij}$,$ N^a_{ij}$,
$ M^{ab}_{ij}$, $ N^{ab}_{ij}$ are their 5-vector and 10-tensor
channel counterparts respectively,
and $g_{sbd},g_{vbd},g_{tbd},g_{scur},g_{vcur},g_{tcur}$ are the
coupling constants in the corresponding channels.
Again the site $i$ and $j$ forming the bond $\avg{ij}$
can be at an arbitrary distance apart.
The bond interactions can be decoupled by introducing the HS
field in each channel respectively.
Following the same reasoning as the case of the spin 1/2,
the bond-interactions keep the fermion determinant positive definite
provided all these coupling constant non-negative.
Similarly, with $g_{sbd}, g_{vbd}, g_{tcur}>0$, the algorithm can be
applied to study the singlet, quintet bond orders and the 10-fold current
order, while with $g_{scur}, g_{vcur}, g_{tbd}>0$,  it is not useful for
study applied to study the singlet, quintet current  orders
and the 10-fold bond order.

\section{Conclusion}\label{sect:conclusion}
The sign problem of the fermionic QMC algorithm is one of the most
important problems in theoretical physics. Its solution would
practically give an universal computational method to solve models
with strong correlations. 
The rigorous theorem established in this work shows that the minus 
sign problem can be eliminated for a much wider class of models
than before,  in which the fermion matrix is invariant under an 
anti-unitary symmetry similar to the
time reversal symmetry in quantum mechanics. The method of $T$-invariant
decomposition does not only provide a deep connection
between the sign problem and the time reversal symmetry, it also
leads to practical algorithms which can be applied to many
interesting models with strong correlations.
Using this algorithm, a new class of models with strong
correlation can be simulated, and some novel and exotic ground
states have been firmly established.

We conclude this paper with an optimistic outlook. Even though our
method can only be applied presently to models with definite
constraints among the interaction parameters, we believe that the
deep symmetry connections revealed in this work could guide us in
future works, and might eventually lead to the complete
elimination of the sign problem.

\vspace{3mm}
{\it Note added in proof:}\\
After the paper has been accepted, we learned that a similar version of
the theorem of T-invariant decomposition had been discussed in the context
of lattice gauge theory \cite{hands2000}. However, they did not consider the
case that I+B is not diagonalizable. Our proof is valid regardless of
whether I+B is diagonalizable or not, thus is more complete." 

\begin{acknowledgments}
We thank Drs. B. A. Bernevig, S. Capponi, D. Ceperley,  S. Chandrasekharan,
J. P. Hu, D. Scalapino, and T. Xiang for helpful discussions.
This work is supported by the NSF under grant
numbers DMR-0342832 and the US Department of Energy, Office of
Basic Energy Sciences under contract DE-AC03-76SF00515. CW is also
supported by the Stanford Graduate Fellowship program.
\end{acknowledgments}

\appendix

\section{$Sp(2n)$ algebra in the  spin $s=n-1/2$ fermion system}\label{spgroup}

We give a brief introduction to the $Sp(2n)$ algebra here.
The $2n$-dimensional Hilbert space on each site can be arranged as a
direct product between a $n$-dimensional and a 2-dimensional space.
The complete basis of eigenstates of $S_z$ are labeled in the sequence of
$| 1\rangle= | n-\frac{1}{2}\rangle,
|2\rangle= | -n+\frac{3}{2}\rangle, ... ,
|n\rangle= | \frac{(-)^{n-1}  }{2}\rangle$,
and
$|\bar 1\rangle=| -n+\frac{1}{2}\rangle,
|\bar 2\rangle= | n-\frac{3}{2}\rangle, ... ,
|\bar n\rangle=| \frac{(-)^{n}  }{2}\rangle$.
The $Sp(2n)$ spinor is defined as
\bea\label{spinor}
\psi=(c_{n-\frac{1}{2}},
c_{-n+\frac{1}{2}}, c_{-n+\frac{3}{2}}, c_{n-\frac{3}{2}}, ....)^T.
\eea
Group elements of $Sp(2n)$ include any $2n\times 2n$ unitary matrix $U$
satisfying $U^T R U=R$ or equivalently $ R^{-1} U R =U^*$
\cite{Hamermesh1989} with the $R$-matrix
\bea
R= I_n \otimes (-i\sigma_2).
\eea
The $R$ matrix is a straightforward generalization of the $R=-i\sigma_2$
in the spin 1/2 case, which also satisfies $R^T=R^{-1}=R^\dagger=-R$.
Clearly, the $Sp(2n)$ group is a subgroup of the $SU(2n)$ group defined
in the $2n$ dimensional space.

In the particle-hole channel, there are $4 n^2$ independent
bilinear operators as $\psi^\dagger_\alpha \psi_\beta
(\alpha=1,..., 2n, \beta=1,..., 2n)$. Among them, the particle
density operator $n=\psi^\dagger_\alpha \psi_\alpha$ is a singlet
under both the $SU(2n)$ and the $Sp(2n)$ group. The time reversal
transformation $T= C R$ is defined as usual, and it satisfies
$T^2=-1$. The other $4n^2-1$ bilinear operators form the
generators (adjoint representation) for the $SU(2n)$ group. They
can be decomposed into two classes according to their
transformation properties under the $T$ operation. The first class
contains $n (2n+1)$ elements which forms the generators of the
$Sp(2n)$ group as denoted as $ \psi^\dagger_\alpha
X^b_{\alpha\beta} \psi_\beta ( b=1\sim 2 n^2 +n )$. $X^b$ can be
expressed in terms a direct product between the $SU(n)$ and
$SU(2)$ generators. We define the $SU(n)$ generators as \bea
&&(M^{(1)}_{ij})_{lk}=\frac{1}{2}(\delta_{il}\delta_{jk}
+\delta_{ik}\delta_{jl})~~ (1\le i<j\le n), \nonumber \\
&&(M^{(2)}_{ij})_{lk}=\frac{-i}{2}(\delta_{il}\delta_{jk}
-\delta_{ik}\delta_{jl})~~ (1\le i<j\le n)  , \nonumber \\
&&M^{(3)}_j=\frac{\mbox{diag} (1,..., 1, -(j-1), 0,...,0)}{\sqrt{2 j(j-1)}}
~~
(2\le j\le n), \nonumber \\
\eea where $M_{ij}^{(1)}, M_{ij}^{(2)},M_{j}^{(3)}$  are the
$n\times n$ dimensional generalization of the $SU(2)$ Pauli
matrices $\sigma_{x,y,z}$ respectively. Counting the numbers of
$SU(2n)$ generators, there are $n(n-1)/2$ real symmetric
$M_{ij}^1$'s, $n(n-1)/2$ imaginary anti-symmetric $M_{ij}^2$'s,
and $n-1$ real diagonal $M_{ij}^3$'s. Then the $Sp(2n)$ generators
$X^b$ can be expressed as
\bea
    M_{ij}^{(2)} \otimes I_2, ~~  M_{ij}^{(1)}\otimes  \vec \sigma,
~~  M_j^{(3)}\otimes \vec \sigma, ~~ I_n \otimes \vec \sigma,
\eea
Because $M_{ij}^{(1)}$'s and $M_{l}^{(3)}$'s are real and $M_{ij}^{(2)}$'s
are purely imaginary, the $Sp(2n)$ generators are odd under the time
reversal: $T^{-1} X^b T= - X^b$.
The second class bilinears $\psi^\dagger_\alpha
Y^a_{\alpha\beta} \psi_\beta $ have $2n^2-n-1$ elements.
$Y^a (a=1,..., 2n^2-n-1)$ are given by
\bea
M_{ij}^{(2)}\otimes \sigma_i
~~M_{j}^{(1)} \otimes I_2,
~~M_j^{(3)}\otimes I_2,
\eea
which are even under the time reversal: $T^{-1} ~Y^a ~T =Y^a$.

These $4n^2$ bilinear operators are not independent of each other, but are
related by the Fierz identity.
The total Hilbert space for one site has the dimension of $2^{2n}$,
which can be decomposed into subspaces with different particle number
$r(0\le r \le 2n)$.
Each of them form the totally anti-symmetric representation of the $SU(2n)$
group $1^r$.
The Casimir value of the $SU(2n)$ group in such representations are
$r(2n+1) (2n-r)/(2n)$.
Thus we arrive the Fierz identity for the spin $n-\frac{1}{2}$ system as
\bea
&& \sum_b (\psi^\dagger_{i\alpha} X^{b}_{\alpha\beta} \psi_{i\beta})^2
+  \sum_a (\psi^\dagger_{i\alpha} Y^{a}_{\alpha\beta} \psi_{i\beta})^2
\nonumber \\
&+& \frac{2n+1}{2n} (\psi^\dagger_\alpha \psi_\alpha-n)^2
=\frac{2n^2+n}{2}.
\eea

The onsite pairing operators can be easily formed by using the $R$ matrix.
Due to the Pauli's exclusion principle, the total spin for a s-wave pair
can only be $0,2, ..., 2n-2$.
The singlet pair operator is also the $Sp(2n)$ singlet operator.
It can be written as $\psi^\dagger_\alpha
R_{\alpha\beta} \psi^\dagger_\beta$, which was studied extensively
in a $Sp(2n)$ generalization of the Heisenberg antiferromagnet
\cite{sachdev1991}.
The other $2n^2-n-1$ pairing operators with total spin $2,4, ..., 2n-2$
together form a representation of $Sp(2n)$
as $\psi^\dagger_\alpha (i) (RY^a)_{\alpha\beta} \psi^\dagger_\beta(i)$.
When all the interaction parameters are equal, these $n(2n-1)$ pairing
operators together form anti-symmetric  representation of $SU(2n)$ of
$1^r$ (r=2).

\end{document}